\documentclass[journal,twocolumn]{IEEEtran}
\title{Secure Massive IoT Using Hierarchical Fast Blind Deconvolution}

\author{
\IEEEauthorblockN{Gerhard Wunder\IEEEauthorrefmark{1},   \and Ingo Roth\IEEEauthorrefmark{2},   \and Rick Fritschek\IEEEauthorrefmark{3},   \and Benedikt Gro{\ss}\IEEEauthorrefmark{4},   \and Jens Eisert\IEEEauthorrefmark{5},  }
\IEEEauthorblockA{ \\Freie Universit\"at Berlin \\
Email: \{\IEEEauthorrefmark{1}g.wunder,
\IEEEauthorrefmark{2}i.roth,
\IEEEauthorrefmark{3}rick.fritschek,
\IEEEauthorrefmark{4}benedikt.gross\}@fu-berlin.de,
\IEEEauthorrefmark{5}jenseisert@gmail.com
 }
}

\usepackage[utf8]{inputenc}
\usepackage{amsmath,amssymb,amsthm,amsfonts}
\usepackage{algorithm}
\usepackage{algorithmic}
\usepackage{graphicx}
\usepackage{enumerate}
\usepackage{stackengine}
\usepackage{color}
\usepackage{csquotes}

\usepackage{todonotes}
\usepackage{mathtools}
\usepackage{manfnt}
\usepackage{balance}

\newcommand{\R}{\mathbb{R}}
\newcommand{\C}{\mathbb{C}}
\newcommand{\N}{\mathbb{N}}

\let\vec\relax
\DeclareMathOperator{\vec}{vec}
\newcommand{\norm}[1]{\left\Vert#1\right\Vert}
\newcommand{\abs}[1]{|#1|}
\DeclareMathOperator*{\argmin}{argmin}
\DeclareMathOperator*{\argmax}{argmax}
\DeclareMathOperator{\supp}{supp}
\newcommand{\st}{\text{ s.t. }}
\newcommand{\A}{\mathcal{A}}
\newcommand{\T}{\mathcal{T}}
\renewcommand{\S}{\mathcal{S}}
\renewcommand{\mod}{\operatorname{mod}}
\newcommand{\circulant}{\operatorname{circ}}
\newcommand{\cconv}{\circledast}

\stackMath

\providecommand{\U}[1]{\protect\rule{.1in}{.1in}}
\providecommand{\U}[1]{\protect\rule{.1in}{.1in}}
\definecolor{ingo}{RGB}{139,0,0}

\definecolor{gerhard}{RGB}{0,0,139}

\definecolor{rick}{RGB}{0,139,0}

\definecolor{bene}{RGB}{10,120,60}

\begin{document}

\maketitle
\begin{abstract}
The Internet of Things and specifically the Tactile Internet give rise to significant challenges for notions of security. In this work, we introduce a novel concept for secure massive access. The core of our approach is a fast and low-complexity blind deconvolution algorithm exploring a bi-linear and hierarchical compressed sensing framework. We show that blind deconvolution has two appealing features: 1) There is no need to coordinate the pilot signals, so even in the case of collisions in user activity, the information messages can be resolved. 2) Since all the individual channels are recovered in parallel, and by assumed channel reciprocity, the measured channel entropy serves as a common secret and is used as an encryption key for each user. We will outline the basic concepts underlying the approach and describe the blind deconvolution algorithm in detail. Eventually, simulations demonstrate the ability of the algorithm to recover both channel and message. They also exhibit the inherent trade-offs of the scheme between economical recovery and secret capacity.
\end{abstract}

\section{Introduction}
\setcounter{page}{1} 
Over the last decades, major developments in communication technologies 
have radically altered the way we communicate. This entails difficult network challenges from the technological side. As the sheer volume 
of data being transmitted is growing, these challenges are concomitant with
new demands on the security of the communication channels. To accompany
the significant challenges of security of communication in the realm of big data, 
novel physical layers of security will have to be identified and developed.
This seems particularly relevant in the context of the 
Internet of Things (IoT) and the Tactile Internet (TI).
In this work we will show that sparse signal processing can be incorporated
naturally within the concept of massive IoT, 
including the TI and embedded security. We go on to demonstrate that it indeed exhibits a new degree of freedom in the design of (low-complexity) algorithms,
naturally entailing new interesting trade-offs such as compressibility versus
secrecy \cite{5g2017_JSAC,Wunder2015_ACCESS}.

Our specific innovations are as follows: We propose a \emph{fast},
\emph{scalable}, and \emph{secure} access procedure with low complexity
\cite{5g2017_JSAC,Wunder2015_ACCESS}. At the heart of our approach is a new fast
blind deconvolution algorithm based on bilinear compressed sensing (CS)
and hierarchical sparsity frameworks \cite{Wunder2017_ASILOMAR,RothEtAl:2016,Wunder2015_ASILOMAR,Wunder2015_ARXIV}.
The proposed algorithm has the additional advantageous feature of being inherent to low-complexity by avoiding
semi-definite programming techniques. Using blind deconvolution for the uncoordinated massive access
has two appealing features: 
\begin{enumerate}[i)]
\item There is no need to coordinate the pilot signals, so even in case of
collisions user activity and information messages can be resolved.
\item Since all the individual channels can be recovered in parallel, and by assumed channel reciprocity,
the measured channel entropy serves as a common secret and is used as an encryption key for each user \cite{Zenger2014_SIOT}.
\end{enumerate}
In this work, we will  outline the underlying basic concepts, and describe the proposed blind deconvolution algorithm in
detail. Eventually, simulations demonstrate the (not at all obvious) ability of the algorithm to
recover both channel and message, and also nicely reveal the inherent trade-offs.  If a channel is sparser, the recovery is improved but at the same time less entropy for key generation is available. Hence, while the recovery can be achieved more economically, the secrecy properties are degraded. \par
\textbf{Basic notations}
\begin{itemize}
\item The circular convolution of two vectors $f,g\in\C^n$ will be denoted by $f\cconv g$ and is defined as 
\begin{equation}
\label{def:circular_convolution}
(f\cconv g)_j:=\sum\limits_{i=1}^n f_jg_{(i-j+1)\mod n}.
\end{equation}
\item $\norm{\cdot}$ will denote either the $\ell_2$-norm of a vector or the Frobenius-norm of a matrix depending on the context. 
\item For a set $\S$ let $\abs{\S}$ denote its cardinality. For any positive $N\in\N$ we define $[N] \coloneqq \{1,2,\ldots,N\}$.
\item For a vector $x\in\C^n$ we denote by $\abs{\cdot}_0$ the function that returns the number of non-zero elements of $x$, i.e. 
\begin{equation}
\label{def:counting_norm}
\abs{x}_0 := \abs{\{i\in[n] : x_i\neq 0\}}.
\end{equation}
\item The transpose/Hermitian of a matrix $A$ with complex entries will be denoted $A^T$ and $A^H$, respectively.
\item The Kronecker product of the matrices $A\in\C^{M\times N}$ and $B\in\C^{O\times P}$ is denoted by $A\otimes B$ and is defined as the $\C^{MO\times NP}$ block matrix
\begin{equation}
\label{def:kronecker_product}
A\otimes B := \begin{bmatrix}
a_{1,1}B & a_{1,2}B &\hdots &a_{1,N}B \\
a_{2,1}B & \ddots & \ddots & \vdots \\
\vdots & \ddots & \ddots& a_{M-1,N}B \\
a_{M,1}B & \hdots & a_{M,N-1}B & a_{M,N}B
\end{bmatrix},\end{equation}
where $a_{i,j}$ is the $(i,j)$-entry of $A$.
\item The map $\vec: \C^{M \times N} \to \C^{MN}$ is the column-wise vectorization of a matrix, i.e.\ it stacks the columns of a matrix into a long vector.
\item $\circulant(v)$ denotes the circulant matrix of a vector $v\in\C^n$, which is defined as
\begin{equation}
\label{def:circulant_matrix}
\circulant(v) := \begin{bmatrix}
v_1 & v_{n-1} & \hdots & v_3 & v_2 \\
v_2 & v_1 & v_{n-1} & & v_3 \\
\vdots & v_2 & v_1 & \ddots & \vdots \\
v_{n-2} & & \ddots & \ddots & v_{n-1} \\
v_{n-1} & v_{n-2} & \hdots & v_2 & v_1 
\end{bmatrix}.
\end{equation}
\end{itemize}

\section{System model}
We consider a secure random access scenario where access point \enquote{Alice}  with $N_{t}$ antennas communicates with $N_{r}$ \enquote{Bobs}, which are low-complexity devices, equipped with a single antenna each. Furthermore, we assume an OFDM signal
model, so that essentially all wireless channel operations become cyclic, acting by the $\circulant(\cdot)$ operation. The
communication is bi-directional and TDD in $T\geq 1$ time slots $t_{0},t_{1},\dots,t_{T-1}$ in the following fashion:
\begin{itemize}
\item First, Alice sends out multiple beacon OFDM symbols so that the
Bobs can synchronize and measure the channels to each of Alice's antennas.
From the measured channels each Bob generates a key and encrypts its message.

\item Subsequently all Bobs transmit in an uncoordinated fashion their
encrypted messages in the same slot while no pilot signaling is used. Alice
uses a blind deconvolution algorithm to simultaneously estimate
the channels \textit{and} the signals \enquote{in one shot}.   
\end{itemize}

\subsection{Wireless channel properties}
The most important random entity is the wireless channel from Alice to all the
Bobs and from the Bobs to Alice per antenna. We use the following convention for the
bi-directional communication: $p$ is the index of the transmitting antenna, $q$ of
the receiving antenna, and $i$ represents the delay domain in some time
slot. Hence, the matrix $H_{p}^{a\rightarrow b}=(h_{p,q,i}^{a\rightarrow b})$ that represents the
wireless channels from Alice's $p$-th antenna to \textit{all} Bobs is given by%
\begin{equation}
H_{p}^{a\rightarrow b}=\left[
\begin{array}
[c]{lllll}%
\vdots &  & h_{p,q,1}^{a\rightarrow b} &  & \vdots\\
h_{p,1,i}^{a\rightarrow b} & \dots & \vdots & \dots & h_{p,N_{r},i}^{a\rightarrow b}\\
\vdots &  & h_{p,q,N_{d}}^{a\rightarrow b} &  & \vdots
\end{array}
\right]  \in\C^{N_{d}\times N_{r}}
\end{equation}
for $p=1,\dots,N_{t}$. In addition, the matrices $H_{p}^{b\rightarrow a}=(h_{pqi}^{b\rightarrow a})$
representing the wireless channels from $p$th Bob to Alice are given by%
\begin{equation}
H_{p}^{b\rightarrow a}=\left[
\begin{array}
[c]{lllll}%
\vdots &  & h_{p,q,1}^{b\rightarrow a} &  & \vdots\\
h_{p,1,i}^{b\rightarrow a} & \dots & \vdots & \dots & h_{p,N_{t},i}^{b\rightarrow a}\\
\vdots &  & h_{p,q,N_{d}}^{b\rightarrow a} &  & \vdots
\end{array}
\right]  \in\C^{N_{d}\times N_{t}}
\end{equation}
for $p=1,\dots,N_{r}$. Notably, we impose a typical structural
assumption for wireless channels: Each column vector $h_{p,q}=(h_{p,q,i})$
contains only $N_{d}\ll N$ coefficients, where $N_{d}$ is called delay spread
of the Channel Impulse Response (CIR) of any $p$-th/$q$-th pair that gets
transmitted/received. This is a common assumption, e.g.\ for OFDM systems.

Now, the received time-space signal (represented by rows and columns,
respectively) in some time slot for Alice is given by $Y^{a}\left(
t_{i}\right)  \in\C^{N\times N_{t}}$ and for all the Bobs by
$Y^{b}\left(  t_{i}\right)  \in\C^{N\times N_{r}}$, where $N\gg1$ is
the signal space dimension and $N_{t}$ and $N_{r}$ are the numbers of antennas
that transmit and receive. We assume that the channel coherence time is
essentially larger than the slot time and shall henceforth drop the dependency on the time slot to ease the notation. On each transmit antenna $p$ with
$1\leq p\leq N_{t}$, 
both some \emph{known} and some \emph{unknown}
transmitted sequences $s_{p}^{a/b},x_{p}^{a/b}\in\C^{N}$ are
broadcast. The signals for Alice and Bob in one time slot then become
\begin{subequations}
\label{eqn:sys}%
\begin{align}
\text{Alice} &  \rightarrow\text{Bob: }Y^{b}  =\sum
_{p=1}^{N_{t}}\left(  S_{p}^{a}+X_{p}^{a}\right)  H_{p}^{a\rightarrow b}+Z^{b},\\
\text{Alice} &  \leftarrow\text{Bob: }Y^{a} =\sum
_{p=1}^{N_{r}}\left(  S_{p}^{b}+X_{p}^{b}\right)  H_{p}^{b\rightarrow a}+Z^{a}.%
\end{align}
Here, $S_{p}^{a/b}=\circulant(s_{p}^{a/b}),~X_{p}=\circulant(x_{p}^{a/b}%
)\in\C^{N\times N_{d}}$ are the circulant matrices of the transmitted
sequences as defined in \eqref{def:circulant_matrix}. The matrices $Z^{a/b}$ denote additive white Gaussian
noise with variance $\eta^{2}$.
We will impose the following structural properties:
\end{subequations}
\begin{itemize}
\item \textbf{Reciprocity property}: If not stated otherwise, we assume the
reciprocity property, i.e., if we change the roles of the transmitting antenna
$p$ and the receiving antenna $q$, the channel coefficients are conjugate
complex, i.e., $h_{p,q,i}^{a\rightarrow b}=\left(  {h}_{q,p,i}^{b\rightarrow a}\right)^{\ast}$. Note that this assumption is by far not unrealistic today, as it is already possible to verify with off-the-shelf WiFi devices \cite{vasisht2016decimeter}.

\item \textbf{Natural structural properties}: We assume that out of the
$N_{d}$ channel coefficients, in each column of $H_{p}^{a\rightarrow b},H_{p}^{b\rightarrow a}$ only
$\sigma>0$ of the CIR coefficients are actually non-zero and the exact positions of the
coefficients within $H_{p}$ are unknown, i.e., the channel is $\sigma$-sparse (in
the canonical base).

\item \textbf{Imposed structural properties}: Our final structural assumption
is that the unknown signals $x_{p}$ are $s$-sparse by design in some
known subspaces with bases $Q_{1},Q_{2},\dots$ such that $x_{p}=Q_{p}b_{p}$, where $b_p$ is a
binary vector with $\abs{b}_0 = s$. The rate delivered by this approach is
\[
R=\frac{1}{N}\log_2\binom{N}{s}\text{[bits]}.%
\]
\end{itemize}
In the sequel, we will propose an algorithm that is able to exploit these structural assumptions to recover both the unknown channels and the unknown signals, given only the superposition of their convolutions. 

\subsection{Inherent security of the scheme}
We briefly describe the information theoretic secrecy stemming from the envisioned scheme. It builds on the reciprocity property of the channel and exploits randomness of the channel gain\footnote{Which is due to fading in the wireless channel.} to generate a key and encrypt the message. We refer to the work \cite{WilsonTse07} for an in-depth analysis regarding the use of channel gains for keys, as that was the first rigorous work on the subject. 

{\bf Phase 1}:
\begin{itemize}
\item Alice sends a predefined pilot signal to all Bobs.
\item Each Bob $q$ can measure the complex-valued channel gains $H_{p,q,i}^{a\rightarrow b}=h_{q,p,i}^{b\rightarrow a}\ \forall p,i$.
\item Each Bob encrypts his message $m$ with $c=f(m,\{h_{p,q,i}^{a\rightarrow b}\})$, effectively using the channel as a source of randomness for key generation.
\end{itemize}

{\bf Phase 2}
\begin{itemize}
\item All the Bobs $p$ send their encrypted cipher texts $c_p$ to Alice in an uncoordinated way.
\item Alice receives the superposition of all the convolutions of the cipher text with the respective channels. Now she has a blind de-mixing/de-convolution problem and receives the cipher-texts and complex-valued channel gain pairs $(H_{p}^{b\rightarrow a},c_p)=(H_{p}^{a\rightarrow b},c_p)\ \forall p,q,i$ of every Bob by using our algorithm.
\item Since Alice knows  $H_{p}^{b\rightarrow a}$, which is the same as $H_{p}^{a\rightarrow b}$ due to reciprocity, she can generate the key herself and decrypt the cipher-texts.
\end{itemize}

We note that small variations between both channels, i.e. small violations of reciprocity do not matter, since we can adjust the key generation process. One can for example quantize the channel gain coarse enough to equalize the keys. This would lower the achievable key rate, but would not impact the security of the scheme, due to the assumed independence between the channel gains from Alice to Eve and Alice to Bob. However, a detailed analysis shall be carried out in follow-up work.

\section{Formulation as blind de-convolution problem}
\subsection{Single user case}

For the purpose of exposition, we first consider the case of a single user and a single antenna. Bob sends the signal $x^b$ over the channel $h^{b\rightarrow a}$ to Alice, who receives
\begin{equation}
\label{eq:conv_single_user}
y^a = h^{b\rightarrow a}\cconv x^b = \circulant(x^b)h^{b\rightarrow a}.
\end{equation}
Using the so-called lifting trick, which was introduced in the context of phase retrieval \cite{Candes2013,Candes22013} and later generalized to blind deconvolution problems \cite{Ahmed2014}, this bi-linear equation can be transformed into a linear one as
\begin{equation}
\label{eq:lifted_bd_single_user}
y^a = B\vec\left(x^b(h^{b\rightarrow a})^T\right)+z^a.
\end{equation}
Here, $B$ is a suitable matrix with $(B)_{i,(j,k)}=\delta
_{i,j+k\mod N}$ ($(j,k)$ is a double index notation), which is
composed as%
\begin{equation}
B=\left(
\begin{array}
[c]{ccccc}%
10...0 & 0...01 & 0...10 & ... & 01...0\\
01...0 & 10...0 & 0...01 & ... & ...\\
... & ... & ... & ... & ...\\
... & ... & ... & ... & ...
\end{array}
\right)  \in\left(  0,1\right)  ^{N\times N^{2}},%
\end{equation}
and $z^a$ is a Gaussian noise vector. The sparse signal model $x^b = Qb^b$ with the random coding matrix $Q\in\C^{N\times E}$ and $s$-sparse binary vector $b^a\in\{-1,1\}^E$ of length $E$ can be incorporated in the formulation to yield
\begin{equation}
\label{eq:model_single_user}
y^a = \underbrace{B(I_{N_d}\otimes Q)}_{=:M}\vec(b^b(h^{b\rightarrow a})^T).
\end{equation}
By this procedure, the blind deconvolution problem of recovering $h^{b\rightarrow a}$ and $x^b$ from the measurement $y^a$ is turned into a matrix recovery problem in $X = b^b(h^{b\rightarrow a})^T$, given the linear measurement operator $\A : \C^{N_d\times E}\rightarrow \C^N$, defined by \eqref{eq:model_single_user}. The factors $h^{b\rightarrow a}$ and $b^b$ can be obtained from $X$ as the first left and right singular vectors of the SVD of $X$.

\subsection{Multi-user case}
In the more general case of multiple Bobs, each of Alice's antennas receives a superposition of signals, each convolved with its respective channel,
\begin{equation}
\label{eq:conv_multi_user}
y_q = \sum\limits_{p=1}^{N_r} h_{p,q}\cconv Q_p b_p +z_q \quad \text{ for }q=1,\ldots,N_t,
\end{equation}
where we have dropped the superscripts indicating the sender and receiver to simplify the notation. The lifting trick can be applied to each summand, resulting in 
\begin{equation}
\label{eq:lifted_bd_multi_user}
y_q = \sum\limits_{p=1}^{N_r} B(I_{N_d}\otimes Q_p)\vec(b_p h_{p,q}^T) + z_q.
\end{equation}
In comparison to \eqref{eq:model_single_user}, this is a (more challenging) problem of simultaneous blind deconvolution and blind de-mixing. Work on this problem has been done in ref.\ \cite{Ling2015}. 
Problem \eqref{eq:lifted_bd_multi_user} can be brought into the form
\begin{equation}
\label{eq:model_multiuser}
y_q = M\vec(X_q)+z_q \quad \text{ for }q=1,\dots,N_t,
\end{equation}
with the big system matrix
\begin{equation}
\label{eq:big_system_matrix}
M = B\begin{bmatrix}
I_{N_d}\otimes Q_1 \\
\vdots\\
I_{N_d}\otimes Q_{N_r}
\end{bmatrix}^T\in\C^{N\times {N_d}\cdot E\cdot N_r},
\end{equation}
and the unknown $X_q = [X_{1,q} X_{2,q} \hdots X_{N_r,q}]$ with $X_{p,q} = b_p h_{p,q}^T$. 
With the structural assumptions that each channel $h_q$ is $\sigma$-sparse, each $b_q$ is $s$-sparse and only $\mu$ of the $N_r$ users are active at a time, 
the vectorization $\vec(X_q)\in\C^{{N_d}\cdot E\cdot N_r}$ becomes a hierarchically $(s,\sigma,\mu)$-sparse vector. 
The final equation for the multi-user, multi-antenna setup is then
\begin{equation}
\label{eq:mu_ma}
\underbrace{\begin{bmatrix}
y_1^T \\ \vdots \\ y_q^T \\ \vdots \\ y_{N_t}^T
\end{bmatrix}^T}_{=Y\in\C^{N\times N_t}} = \underbrace{B\begin{bmatrix}
I_{N_d}\otimes Q_1 \\
\vdots\\
I_{N_d}\otimes Q_{N_r}
\end{bmatrix}^T}_{=M\in\C^{N\times {N_d}\cdot E\cdot N_r}}\cdot\underbrace{\begin{bmatrix}
\vec(X_{1})^T \\ \vdots \\ \vec(X_q)^T \\ \vdots \\ \vec(X_{N_t})^T
\end{bmatrix}^T}_{=X\in\C^{{N_d}\cdot E\cdot N_r\times N_t}}.
\end{equation}
It is worth noting that the columns of $X$ are jointly sparse, since the antennas are close to each other, and hence for each $p$, the channels $h_{p,q}$ have the same support for all $q$.

\section{Fast blind de-convolution algorithm}
\subsection{Prior work}
There exists a number of recent works on solution strategies for the blind deconvolution problem and the extended blind deconvolution and blind de-mixing problem using the different approaches. Convex approaches use the formulation
\begin{equation}
\label{eq:bd_convex}
\min\limits_{X}\varphi(X) \quad \st \A(X) = y,
\end{equation}
where $X$ is the unknown matrix variable, $\A$ is the linear measurement operator and $y$ the given data. The objective function $\varphi(\cdot)$ is used to incorporate structural assumptions on $X$ that can be exploited to find a unique solution to the under-determined system $\A(X)=y$. In ref.\ \cite{Ahmed2014}, the nuclear norm $\varphi(X)=\norm{X}_\ast$ is used, exploiting the fact that $X$ as an outer product of $b$ and $h$ is a rank one matrix. Instead of sparsity priors for $b$ and $h$, in ref.\ \cite{Ahmed2014} the authors assume that both vectors are in known low-dimensional subspaces. This setting was generalized to include the de-mixing of multiple convolutions in ref.\ \cite{Ling2015,JungEtAl:2017}. 

To relax the subspace assumption to sparse vectors, it seems natural to linearly combine the regularizers promoting low-rankness and sparsity of the matrix, i.e. $\varphi(X) = \norm{X}_\ast + \lambda \norm{X}_1$. But in fact one can show that the linear combination does not yield an improved sampling complexity, compared to just using one of the regularizers \cite{OymakEtAl:2015}. Furthermore,  convex formulations including the nuclear norm are semidefinite programs and can be solved by popular interior-point solvers such as SDPT3 \cite{toh1999sdpt3} or SeDuMi \cite{sturm1999using}. These SDP-solvers have the drawback of being prohibitively slow and memory consuming for large scale problems, as their computational and storage complexity typically scales cubically in the system size.

For this reason, subsequent convex approaches focused on exploiting the sparsity of $X$ and structured versions thereof. Ling and Strohmer minimize $\varphi(X)=\norm{X}_1$ assuming that at least one of the factors $h$, $b$ is sparse and hence also $X$. In this setting the sparsity of $X$ is structured since each column is either vanishing or dense. This block-sparse structure motivated the use of the objective function is $\varphi(X)=\norm{X}_{1,2}$, which is defined as the sum of the column norms of $X$, in ref.\ \cite{Flinth2017}. The current work follows in this line of research, further incorporating the sparsity structures inherent to the problem, if both vectors $h$ and $b$ are assumed to be sparse. 

Following a different approach, a number of non-convex algorithms, mostly based on alternating minimization, exist that deal with blind-deconvolution and related problems. For example, the blind deconvolution and blind de-mixing problem, where low-dimensional subspaces for both vectors are known, is tackled in \cite{Li2016a,Ling2017} and the sparse setting is handled in \cite{LeeEtAl:2015}. For these to work properly, a good initial guess for the unknown factors of $X$ is crucial. Therefore, in \cite{Li2016a} a basin of attraction is constructed, and a spectral method is used to obtain an initialization close to the solution. The algorithm of \cite{LeeEtAl:2015} uses a hard thresholding algorithm to compute a sufficiently close initial guess and only then proceeds with their alternating minimization algorithm. This algorithm, however, involves the projection onto a complicated, non-convex set whose success can not be guaranteed.

\subsection{Proposed algorithm}
Motivated by the application in mMTC, the recovery of hierarchical sparse signals from linear measurements was studied in ref.\ \cite{RothEtAl:2016}.
In this work, the HiHTP algorithm was extended to solve the outlined 3-dimensional problem
\begin{equation}
\label{eq:problem_multiuser}
\min\limits_{z\in\C^{{N_d}\cdot E\cdot N_r}}\frac12\norm{y-Mz}^2 \st z \text{ is hierarchically $(s,\sigma,\mu)$-sparse.}
\end{equation}
A hierarchically sparse vector $z\in\C^{{N_d}\cdot E\cdot N_r}$ has the following structure. 
\begin{equation}
\label{eq:hi_sparse}
z = (z^1, z^2, \ldots  \stackunder{z^r}{\overbrace{(z^r_1, z^r_2, \ldots, \stackunder{z^r_j}{\overbrace{(z^r_{1,j}, z^r_{2,j}, \ldots,  z^r_{i,j}, \ldots,  z^r_{D,j})}}, \ldots, z^r_E)}}, \ldots,  z^{N_r})^T
\end{equation}
As described above, only $\mu$ of the vectors $z^r$ are different from zero, or \enquote{active}. The active vectors only have $\sigma$ non-zero blocks, and each of these blocks is $s$-sparse.

HiHTP tries to find such a structured solution to \eqref{eq:problem_multiuser} by repeating the following steps: 
\begin{enumerate}[i)]
\item Perform one gradient step on the current iterate $z^{(k)}$.
\item Determine the support $\S^{(k+1)}$ of the next iterate via hierarchical hard thresholding.
\item Solve a least squares problem on $\S^{(k+1)}$ to obtain the new iterate $z^{(k+1)}$.
\end{enumerate}
The details of each step are explained below.

\textbf{Gradient step:}
The gradient of the objective function from \eqref{eq:problem_multiuser} at $z^{(k)}$ is given by 
$M^T(Mz^{(k)}-y)$.
Hence, the intermediate point is given by 
\begin{equation}
\label{eq:intermediate}
\tilde{z}^{(k+1)} = z^{(k)}+M^T(y-Mz^{(k)}).
\end{equation}
\textbf{Hierarchical hard thresholding:}
In this step, the support of the next iterate is found by thresholding the intermediate point defined in \eqref{eq:intermediate} with the algorithm explained below.
Define the hard thresholding operator $\T_s: \C^n\rightarrow [n]^s$ applied to a vector $g\in\C^n$ as
\begin{equation}
\label{eq:hard_thresholding}
\T_s(g) = \argmax\limits_{\{i_1,\ldots, i_s\}\subset[n]}\sum\limits_{k=1}^s \abs{g_{i_k}}.
\end{equation}
The hierarchical hard thresholding operator with three layers, denoted by $\T_{(s,\sigma,\mu)}$, is given by the following algorithm.

\begin{algorithm}[H]
\caption{Support via hierarchical hard thresholding}
\begin{algorithmic}
\REQUIRE Structured vector $g\in\C^{D\cdot E\cdot N_r}$ as above, sparsity $(s,\sigma,\mu)$
\FOR{$k=1,\ldots ,N_r$}
	\FOR{$j=1,\ldots ,E$}
		\STATE $I_j^k = \T_s(g_j^r)$
        \STATE $v_j^k = \sum\limits_{i\in I_j} \abs{g_{ij}^k}$
    \ENDFOR
    \STATE $J_k = \T_\sigma(v^k)$
    \STATE $u^k = \sum\limits_{j\in J_k} v_j^k$
\ENDFOR
\STATE $K = \T_\mu(u)$
\STATE $\S = \bigcup\limits_{k\in K}\bigcup\limits_{j\in J_k} I_j^k$
\ENSURE $(s,\sigma,\mu)$-sparse support set $\S$
\end{algorithmic}
\end{algorithm}
Hence, the support in step $k+1$ is computed as 
\begin{equation}
\label{eq:support}
\S^{(k+1)} = \T_{(s,\sigma,\mu)}(z^{(k+1)})
\end{equation}
\textbf{Least-squares problem}
The entries of the next iterate $z^{(k+1)}$  are then computed by solving a least squares problem with support constraints, i.e.
\begin{equation}
\label{eq:least_squares_on_support}
z^{(k+1)} = \argmin\limits_{z\in\C^{N_dEN_r}}\left\{\norm{y-Mz}\st \supp(z)\subseteq\S^{(k+1)}  \right\}.
\end{equation}
The algorithm is stopped, if $\S^{(k+1)} = \S^{(k)}$ or the maximum number of iterations is reached. 
The whole algorithm is summarized below.
\begin{algorithm}[H]
\caption{HiHTP - multi-user case}
\begin{algorithmic}
\REQUIRE Measurement matrix $M$; data $y$; channel- , signal- and user-sparsities $s,\sigma,\mu$
\STATE Set $z^{(0)} = 0$, $k=0$
\REPEAT
\STATE Compute support via hierarchical hard thresholding: \\
$\S^{(k+1)} = \T_{s,\sigma,\mu}\left( z^{(k)} + M^H(y-Mz^{(k)}) \right)$
\STATE Compute the corresponding entries by solving the least-squares problem: \\
$z^{(k+1)} = \argmin\limits_{z\in\C^{N_dEN_r}}\left\{\norm{y-Mz} \st \supp(z)\subseteq \S^{(k+1)}\right\}$ and set $k=k+1$
\UNTIL{stopping criterion is met}
\ENSURE Hierarchical sparse solution $z^*$
\end{algorithmic}
\end{algorithm}

\section{Simulations}
To test the efficiency of the HiHTP-algorithm in the multiuser setting, the following tests were conducted:
We assume for simplicity that Alice only consists of one antenna and that there are $N_r$ Bobs, from which only $\mu<N_r$ are active. The multi-antenna setting will offer further possibilities to improve the performance, since the correlations between the antennas will introduce more structure into the model. The completion of this model and the design of an efficient algorithm for it is currently investigated by the authors. 
For each of the $N_r$ users a $\sigma$-sparse channel $h_k\in \R^{N_d}$ was drawn with the locations of the non-zeros distributed uniformly and entries drawn from the standard normal distribution. The signals were computed as $x_k = Q_k b_k$ were $Q\in \R^{N\times E}$ is a random matrix with entries $Q_{i,j} \sim \mathcal{N}(0,1)$ and $b\in \R^E$ is $s$-sparse with values in $\{-1,1\}$ if the user is active, and $0$ if the user is not active. 
This results in the data $y\in\R^N$,
\begin{equation}
\label{eq:data_multiuser}
y = \sum\limits_{k=1}^{N_r} h_k \cconv Q_k b_k.
\end{equation}
The measurement matrix $M\in\R^{\times {N_d}\cdot E\cdot N_r}$ is such that 
\begin{equation}
\label{eq:measurement_equation}
y = M\vec(X),
\end{equation}
with $X=[b_1h_1^T \ldots b_{N_r}h_{N_r}^T]$.
The experiments were conducted with $N=1024$, ${N_d}=E=128$ and $N_r=10$. The number of active users varied from 2 to 5 and the sparsity levels of $h$ and $b$ varied from 2 to 15. An experiment was classified successful, if the support of $X$ was recovered correctly and the residual was below $10^{-6}$. 
The graphics below show the rate of successful recovery for varying number of active users, averaged over 20 runs per setup. The x- and\ y-axis show the channel sparsity $\mu$ and\ the signal sparsity $s$, respectively.
\begin{figure}
	\centering
    \begin{minipage}[b]{0.475\textwidth}
    \includegraphics[width=\textwidth]{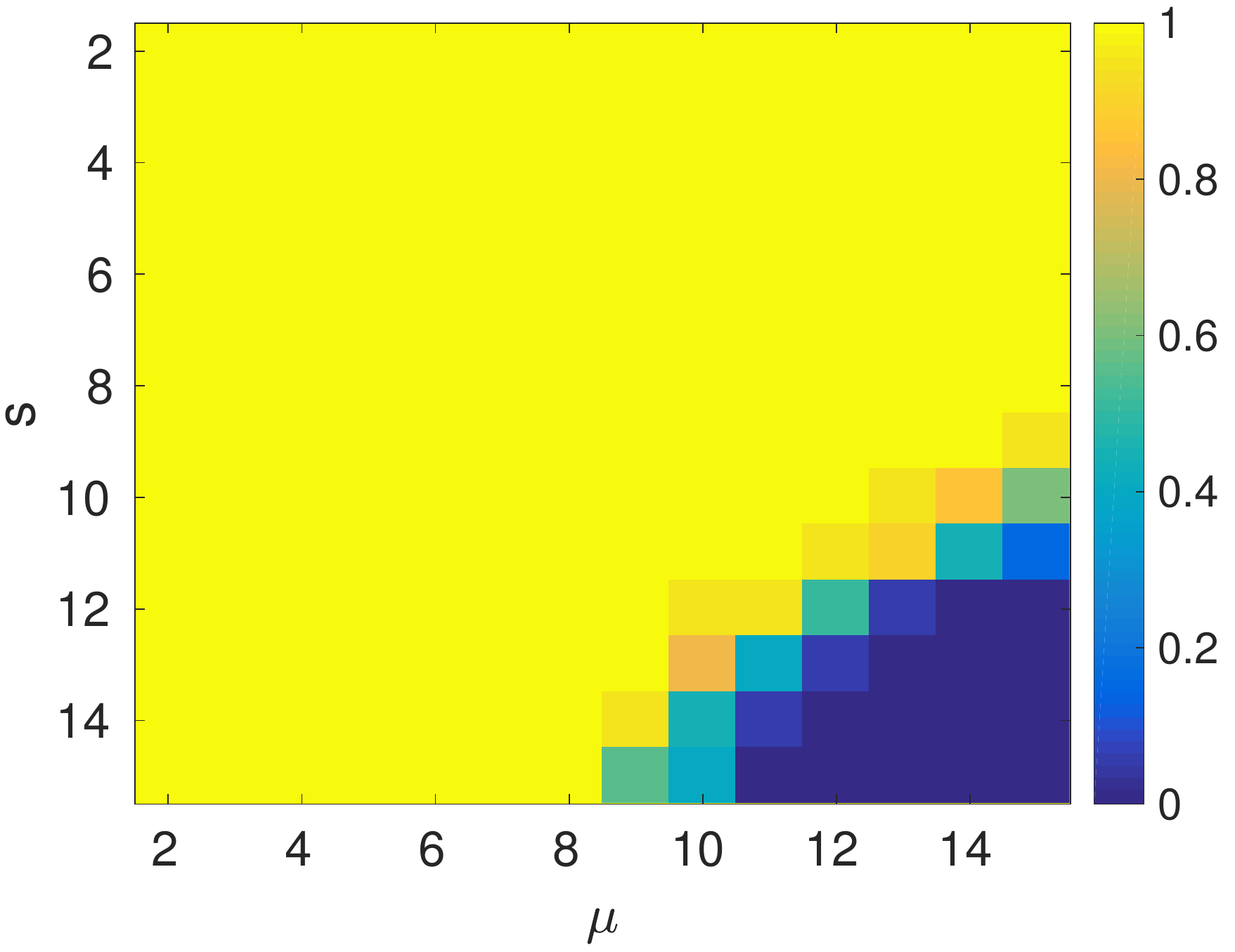}
    \caption{Recovery rate for 2 of 10 active users}
    \end{minipage}
    \hfill
    \begin{minipage}[b]{0.475\textwidth}
    \includegraphics[width=\textwidth]{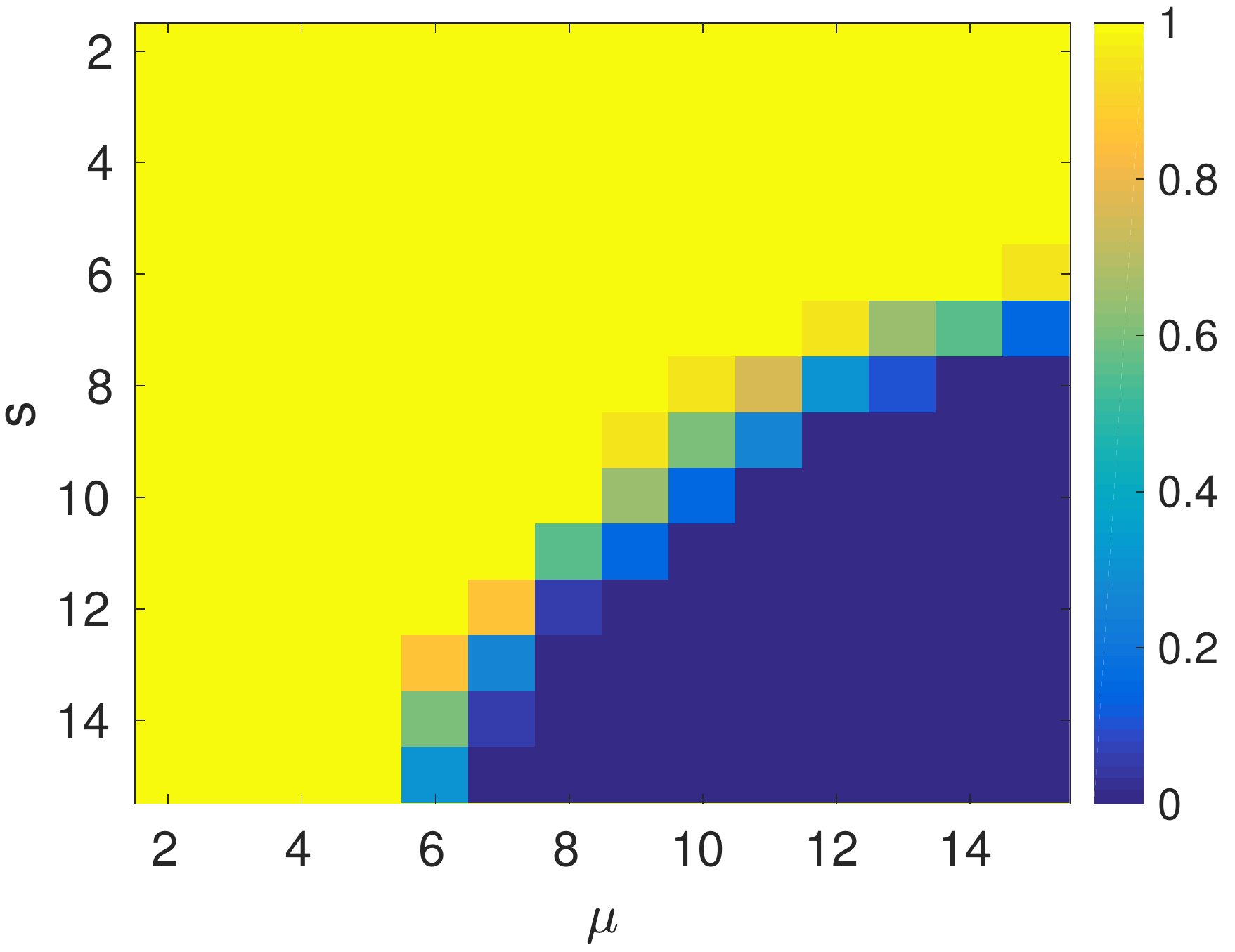}
    \caption{Recovery rate for 3 of 10 active users}
    \end{minipage}
\end{figure}

\begin{figure}
	\centering
    \begin{minipage}[b]{0.475\textwidth}
    \includegraphics[width=\textwidth]{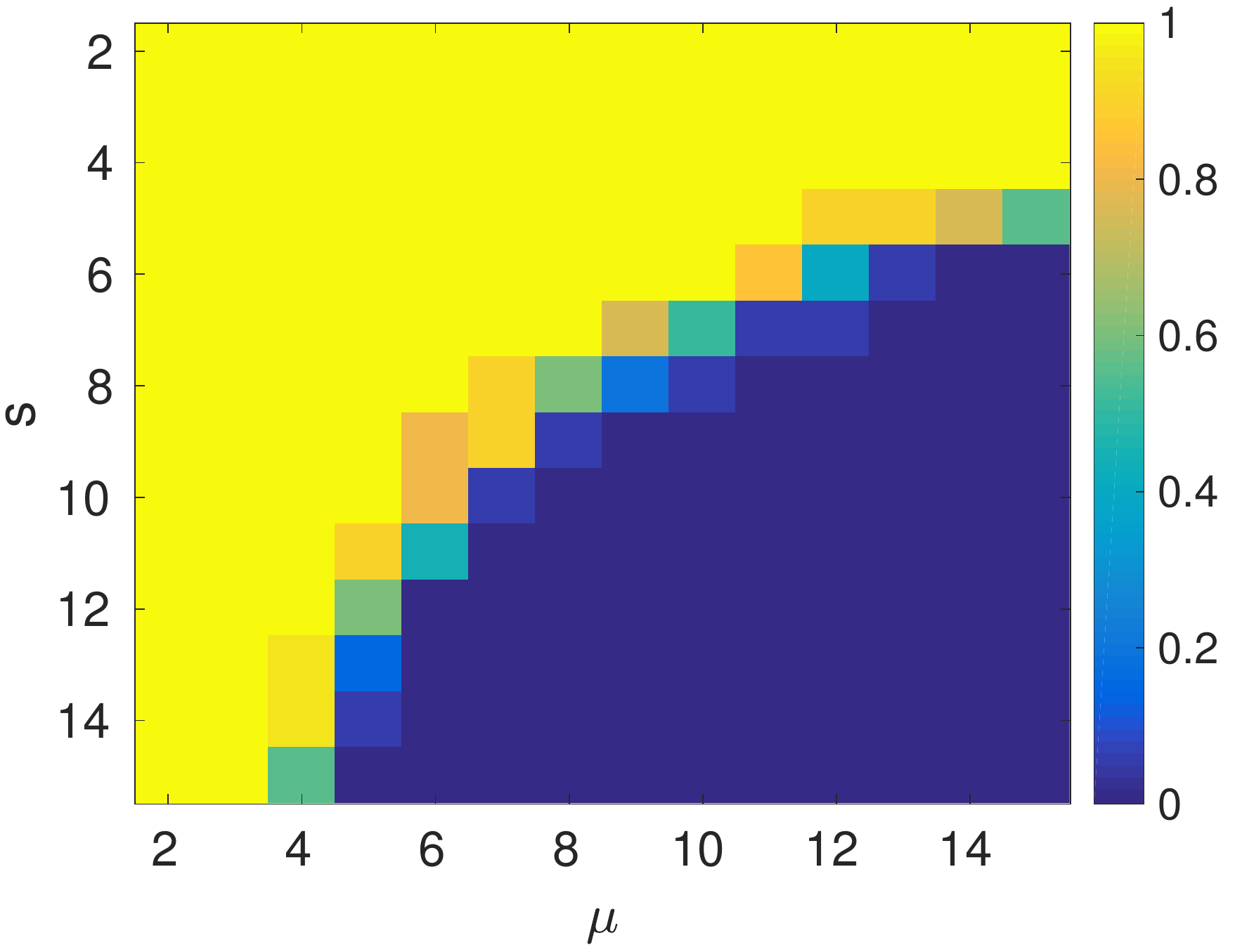}
    \caption{Recovery rate for 4 of 10 active users}
    \end{minipage}
    \hfill
    \begin{minipage}[b]{0.475\textwidth}
    \includegraphics[width=\textwidth]{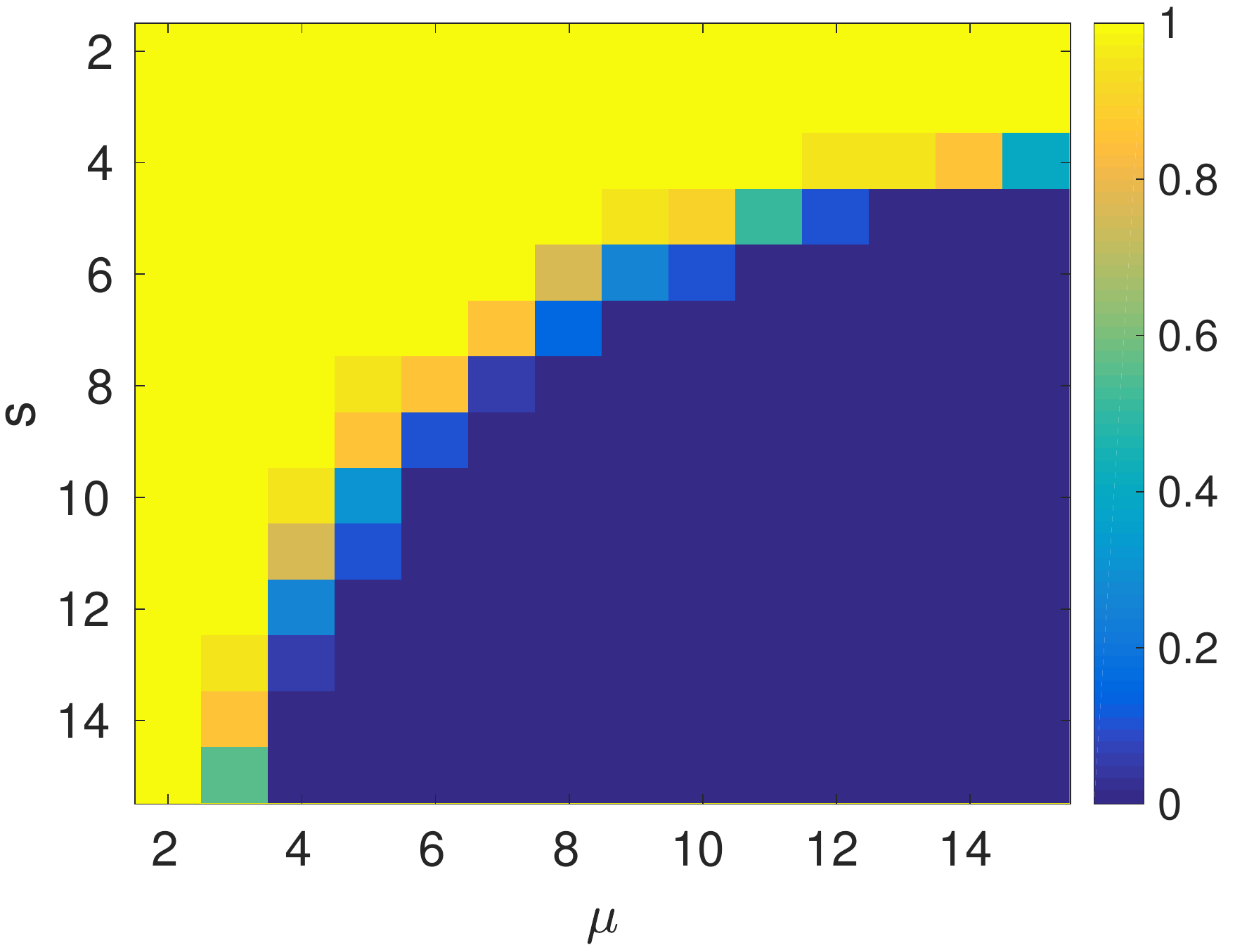}
    \caption{Recovery rate for 5 of 10 active users}
    \end{minipage}
\end{figure}


\section{Conclusions}
We have proposed a new access scheme for IoT applications in which many low complexity devices spontaneously send data to a base station in an uncoordinated fashion and included a physical layer security scheme. The base station is able to recover the signals as well as the channels by employing a fast, scalable blind deconvolution algorithm called HiHTP. The benefit of this novel approach is that it requires no pilot signaling to measure the channels, thus greatly reducing the overhead. This is crucial for next generation wireless communication, where the number of devices will increase dramatically. 
We have provided numerical experiments that show the feasibility of our approach and illustrate the trade-off between the number of active users, the required sparsity of the signals and the channel sparsity. The adaptation of our HiHTP algorithm to the multi-user, multi-antenna case, its robustness to noisy measurements and the proof of rigorous  performance guarantees will be a topic of future research.

\section{Acknowledgements}
We would like to thank the DFG within grants WU 598/7-1, WU 598/8-1, and EI 519/9-1 (DFG
Priority Program on Compressed Sensing) and the Templeton Foundation for support. 
This work has also been performed in the framework of the Horizon 2020 project ONE5G (ICT-760809) receiving funds from
the European Union. The authors would like to acknowledge
the contributions of their colleagues in the project, although
the views expressed in this contribution are those of the authors
and do not necessarily represent the project.

\newpage
\bibliographystyle{IEEEtran}
\bibliography{blind}

\end{document}